# Link updating strategies influence consensus decisions as a function of the direction of communication

Sharaj Kunjar*[1,2], Ariana Strandburg-Peshkin[1,3,4], Helge Giese[4,5,6],
Pranav Minasandra[1,4,7], Sumantra Sarkar[8], Mohit Kumar Jolly[9], Nico Gradwohl[1,4,5]

[1] Department for the Ecology of Animal Societies, Max Planck Institute of Animal Behavior, Konstanz, Germany
[2] Undergraduate Programme, Indian Institute of Science, Bangalore, India
[3] Department of Biology, University of Konstanz, Germany
[4] Centre for the Advanced Study of Collective Behaviour, University of Konstanz, Germany
[5] Department of Psychology, University of Konstanz, Germany
[6] Heisenberg Chair for Medical Risk Literacy and Evidence-based Decisions, Charité – Universitätsmedizin Berlin, Berlin, Germany
[7] International Max Planck Research School for Quantitative Behavior, Ecology and Evolution, Radolfzell, Germany
[8] Department of Physics, Indian Institute of Science, Bangalore, India
[9] Centre for BioSystems Science and Engineering, Indian Institute of Science, Bangalore, India

*Author to whom correspondence should be addressed to: S Kunjar
(sharaj.kunjar@gmail.com)

## Abstract

Consensus decision-making in social groups strongly depends on communication links that determine to whom individuals send, and from whom they receive, information. Here, we ask how consensus decisions are affected by strategic updating of links and how this effect varies with the direction of communication. We quantified the co-evolution of link and opinion dynamics in a large population with binary opinions using mean-field numerical simulations of two voter-like models of opinion dynamics: an Incoming model (where individuals choose who to receive opinions from) and an Outgoing model (where individuals choose who to send opinions to). We show that individuals can bias group-level outcomes in their favor by breaking disagreeing links while receiving opinions (Incoming Model) and retaining disagreeing links while sending opinions (Outgoing Model). Importantly, these biases can help the population avoid stalemates and achieve consensus. However, the role of disagreement avoidance is diluted in the presence of strong preferences – highly stubborn individuals can shape decisions to favor their preferences, giving rise to non-consensus outcomes. We conclude that collectively changing communication structures can bias consensus decisions, as a function of the strength of preferences and the direction of communication.

# Introduction

Decision-making in social groups of humans and non-human animals[1–4] often requires individuals to collectively reach consensus from a set of choices. Animal collectives such as schooling fish or flocking birds must constantly engage in consensus decision-making to maintain group cohesion while moving together[5–9]. Consensus decisions also affect instrumental processes in human societal functioning[10–16], from democratic elections to the establishment and evolution of ideas and norms. Since the advent of information and network technology, online social networks (OSNs) are slowly replacing print and broadcast media as key sources of political communication[17]. By erasing spatial barriers and providing unprecedented access to information (and misinformation), OSNs have become an important platform for opinion exchange, through political debates and public discourse[18–20]. However, OSNs are also prone to structural biases detrimental to democratic decision-making such as information gerrymandering, selective exposure and echo chamber formation[21–27]. Hypothesis regarding causes of such phenomena range from biases in feed algorithms to users selecting self-similar content[24,28,29]. While studies agree on how these processes threaten healthy public discourse and may lead to increased ideological polarization, the mechanisms in play are still under debate[30]. A mechanistic understanding of collective behavior and opinion dynamics in such networks is necessary in order to regulate decision-making and avoid potential pitfalls[31].

Collective decisions emerge as a result of repeated interactions among individuals in a network who express various *opinions*. Individuals may have incentives and beliefs that give rise to *preferences* that shape the opinions they state[32]. When preferences conflict with the collective opinion, individuals may revert or stick to their preferences. The tendency to stick to one's preference can be described as *stubbornness*, which can lead to zealotry and despotic behavior. Previous studies have incorporated such stubbornness through inflexible bots with fixed opinions[33,34], or inherent biases assigned to each individual[35–37]. These studies suggest that highly stubborn individuals can shape collective outcomes to their favor, but the strength of the effect also depends on the underlying network structure. Finally, individuals may also have a bias toward the status quo, in which case they exhibit *inertia*[38,39]. Inertia describes the tendency of an individual to retain their current opinion, irrespective of their preference and available social information.

Within a communication network, individuals can socially influence each other. Opinions spread over links when individuals adopt the opinions of their neighbours or when they convince their neighbours to adopt their opinions. As a result, link structures shape how opinions spread in a network and can facilitate or impede consensus outcomes[16,40,41]. Importantly, communication networks are not always static, and social influence can be determined by individuals' choices of who they communicate with[1,42]. Thus, the strategic selection and updating of links may have a crucial impact on group-level outcomes. As such, manipulating the rules of link selection is a means by which communication network providers may moderate opinion dynamics in their network without interacting with its content.

Traditionally, studies on link updating strategies have focused on homophily (the tendency to seek out similar-minded others) or heterophily (the tendency to seek out dissimilar-minded others)[43–46]. Homophily has been observed in the formation of filter bubbles and echo chambers[24] in social media. Additionally, theoretical studies have shown that heterophily can give rise to complicated network structures[47]. It is therefore known that link formation strategies can affect network structure, which subsequently shapes collective outcomes. However, link formation and dissolution are also often influenced by opinion dynamics, with individuals forming and breaking links based on their observations of whether others agree or disagree with them. This feedback between link updating and opinion dynamics may have important effects on collective outcomes, but its consequences are less well understood. Moreover, homophily and heterophily are often perceived as mutually exclusive properties of networks. Tendencies to seek similar-minded or dissimilar-minded others can co-exist and vary between individuals or subpopulations, and even within each individual over time, though this duality has yet to be incorporated into studies of link updating strategies.

Importantly, the aspect of who has the autonomy to update links in a network is central to how communication structures can be manipulated. In some cases, communication links are updated by the receiver of information, for example, following or muting a public account on Twitter. In others, communication links are controlled by the sender of information, for example, via targeted advertising or setting Twitter circles. When individuals choose who to receive information from, they may personalize their information intake and select who to get influenced by. Similarly, when individuals choose who to send information to, they may personalize their audience and select who to influence. While very few studies[48] have differentiated between user behavior in pulling or pushing information, investigating how the direction of communication interplays with opinion dynamics can yield crucial insights towards the consequences of link updating under different context-specific rules of link selection.

The development of opinions over time has often been investigated via Voter models[49–51]. In these models, agents probabilistically imitate the opinions of others that are connected to them in the network. Recent modifications of the classic Voter model also incorporate link updating behavior. Importantly, the Direct (dVM) and Reverse Voter models (rVM)[52] distinguish between incoming and outgoing directions of communication. Mean-field investigations of voter models have been used extensively in the past to study collective behavior in large groups. Using statistical physics yields valuable insights into social dynamics of large collectives, because qualitative properties of large-scale phenomena often do not depend on the microscopic complexities of a given system but rather on macroscopic symmetries and conservation laws[53]. Examples such as Sznajd[54] and Galam[55] models have also shown the merit of the sociophysics approach by successfully describing voting behavior in the past.

In this study, we examine how opinion dynamics and link updating strategies affect the dynamics of collective decision-making processes, and consequently result in consensus or stalemate outcomes **(Figure 1A)**. We primarily investigate the properties of two models – *Incoming (IM)* and *Outgoing Models (OM)* **(Figure 1B)**. These models distinguish between *incoming* and *outgoing* directions of communication with respect to the individual who can update links. Importantly, incorporating such directionality allows us to distinguish conditions where senders choose who receives their opinions (outgoing) from the conditions where receivers choose who sends them opinions

(incoming). To study the role of link updating strategies on consensus formation, our models consider a probability of link updating that depends on agreement or disagreement. We model *disagreement avoiding* and *agreement avoiding* strategies **(Figure 1C)**, by individuals' probability to break links with disagreeing (different opinion) and agreeing (same opinion) others respectively. Homophily emerges when individuals exhibit high disagreement avoidance and low agreement avoidance tendencies. Similarly, heterophily emerges when individual exhibit high agreement avoidance and low disagreement avoidance. However, unlike homophily and heterophily which are antagonistic by definition, disagreement avoidance and agreement avoidance are independent and can co-exist in an individual's strategy: an individual can be likely to break links with disagreeing others, agreeing others, neither, or both. Finally, we investigate how these link updating strategies interact with how strongly individuals reset to their stable preference (stubbornness) and how strongly they maintain their current opinions (inertia). We use these models to investigate how the co-evolution of link structures and opinions in networks affect collective outcomes.

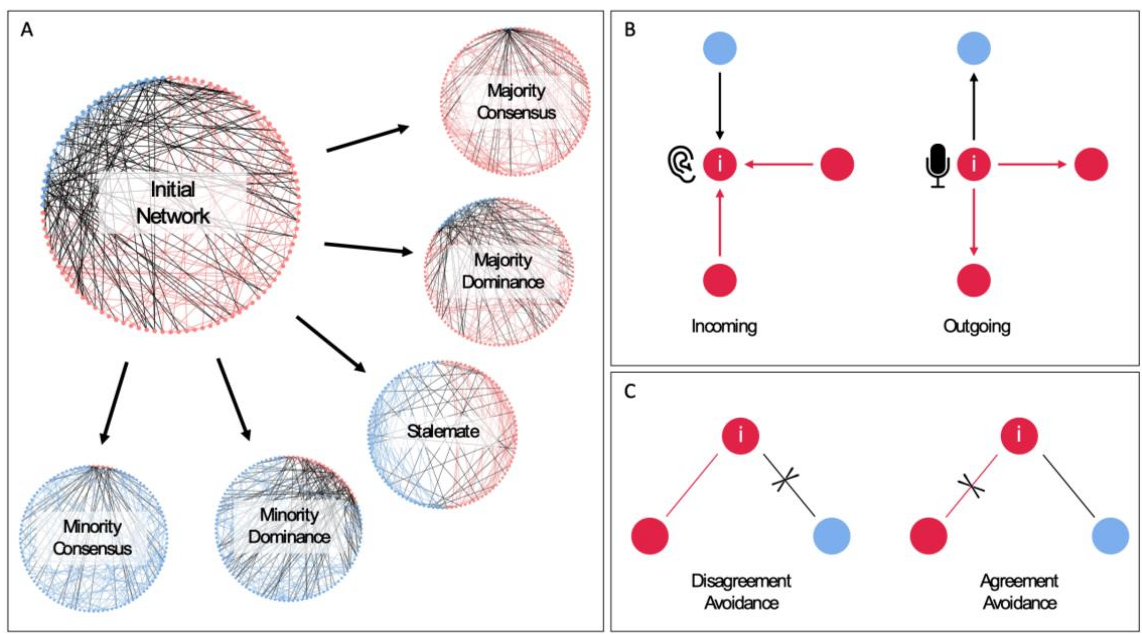

**Figure 1: (A)** Visualization of the communication network of a population with majority (red) and minority (blue) subpopulations. Colors represent the nodes' opinions, and lines represent the links between nodes. Links connecting nodes with the same opinion are shown in the color corresponding to that opinion, whereas links connecting nodes with different opinions are shown in black. Small networks show possible outcomes of the decision-making process starting with this initial configuration, including consensus on majority or minority preference, majority or minority dominance, or stalemate. **(B)** Dependent on the type of model, the link updating individual $i$ may either receive (Incoming model; left) or send (Outgoing model; right) opinions via links. **(C)** For any given link, the focal individual $i$ may avoid disagreement by breaking the link with the blue node (disagreement avoiding strategy; left) or avoid agreement by breaking the link with the red node (agreement avoiding strategy; right).

Using mean-field numerical simulations on the Incoming (IM) and Outgoing (OM) models, we observe how collective outcomes and decision-making speeds vary with individual interactions in the population. Our main question is: how do varying levels of disagreement or agreement avoidance bias the consensus outcome? We also ask how the benefits or detriments of such link updating strategies depend on whether the links are updated by receivers or senders. We identify regions in our parameter space where breaking or retaining links can compensate for the presence of highly stubborn individuals. Beyond that, we examine how behavioral traits such as stubbornness and inertia interplay with link updating strategies to determine group-level decision dynamics in social networks.

# Methods

## 1. Model

We considered a population with $N$ individuals, each holding an opinion: '+' (red in **Figure 2**) or '−' (blue in **Figure 2**) at each point in time, and a time-independent preference for either. The fractions of individuals with '+' and '−' opinions at any time $t$ is given by $n_+(t)$ and $n_-(t)$ respectively, such that $n_+(t) + n_-(t) = 1$. In the beginning (at $t = 0$) each individual starts with an opinion equal to their preference. Therefore, the fraction of individuals with '+' and '−' preferences are $n_+(0)$ and $n_-(0)$. A focal individual $i$ and a non-focal individual $j$ communicate via links, which exist with a population density $l_{ij}$.

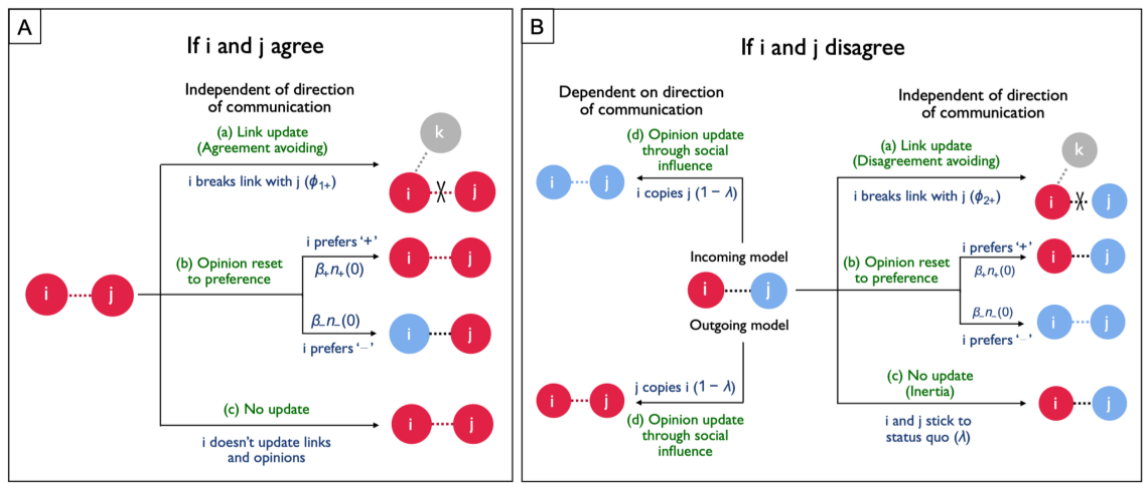

**Figure 2:** The underlying local level processes and decision outcomes of link and opinion updates with i as the sampled focal individual and j as its interaction partner.
If i and j agree **(A),** independent of the direction of communication, individual i does one of three things: **(a)** i avoids agreement, breaking its link with j and creating a new link with a random individual k **(b)** i resets opinion to its preference, irrespective of j's opinion **(c)** i maintains its opinion and keeps the link with j
If i and j disagree **(B)**, the possibilities are: **(a)** i avoids disagreement, breaking its link with j and creating a new link with a random individual k **(b)** i resets opinion to its preference, irrespective of j's opinion. **(c)** i and j maintain their link and opinions due to inertia **(d)** i receives and adopts j's opinion (incoming model) OR i sends and convinces j to adopt i's opinion (outgoing model).

The model (summarized in **Table T1,T2; supplementary**) describes the global outcome of the following process **(Figure 2)**: In each time step, we sample a focal individual $i$ and consider its interaction with a connected individual $j$. Without loss of generality, we assume that '+' denotes the *majority subpopulation* and '−' denotes the *minority subpopulation* in the initial configuration. The co-evolution of links and opinions then occurs through one of the local level interaction processes:

**(a) Link update:** Focal individual $i$ breaks its link with $j$ and makes a link with a random individual $k$. The link updating probability is given by $\phi_{1i}$ if i and j agree and $\phi_{2i}$ if they disagree. Therefore, $\phi_{1i}$ corresponds to the degree of agreement avoidance and $\phi_{2i}$ to the degree of disagreement avoidance among individuals in the respective subpopulations.

(b) **Opinion reset to preference:** Focal individual i ignores social influence and sticks or reverts to its stable preference. The tendency to stick or revert to one's preference is reflected by stubbornness $\beta_+$ and $\beta_-$ for '+' and '−' preferences, respectively.

(c) **No update:** If i and j agree, both links and opinions are conserved. If i and j disagree, they deter social influence by exhibiting inertia and maintaining status quo with probability ($\lambda$), thus again conserving their links and opinions.

(d) **Opinion update through social influence:** If i and j disagree, an opinion update occurs. In the incoming model i adopts j's opinion, whereas in the outgoing model j adopts i's opinion. This update is weighted with a probability $(1 - \lambda)$, where $\lambda$ denotes inertia.

At the population level, the probability of switching to a preference is given by the *net preference*, defined as the product of stubbornness and fraction of population with the given preference: $\beta_+ n_+(0)$ and $\beta_- n_-(0)$ for '+' and '−' preferences respectively. The ratio of net preferences for both the subpopulations is defined as the *preference ratio* $\alpha = \frac{\beta_+ n_+(0)}{\beta_- n_-(0)}$ which indicates the balance of preferences at a population level. Note that all the behavioral parameters (**Table T2; supplementary**) including link updating ($\phi_1, \phi_2$), stubbornness ($\beta_+, \beta_-$) and inertia ($\lambda$) are probabilities and hence range between 0 and 1.

## 2. Mean-field approximation

Firstly, we assume that individuals in the network are unidentifiable, and distinctions are based on their opinion. All nodes with the same opinion are identical and hence, we ignore variance between nodes of the same opinion. The network at any time $t$ is described by the vector $x(t) = \{M(t), l_{+-}(t), l_{++}(t), l_{--}(t)\}$. The four components of the $x$ are:

- The magnetization $M(t) = n_+(t) - n_-(t)$, $M(t) \in [-1, +1]$, is defined as the difference between the fractions of individuals with '+' and '−' opinions respectively. The final state of magnetization $M_{final}$ indicates whether the population reached consensus. We define:

$$\text{Majority consensus: } M_{final} > 0.8$$
$$\text{Majority dominance: } 0.8 \geq M_{final} > 0$$
$$\text{Stalemate: } M_{final} = 0$$
$$\text{Minority dominance: } 0 > M_{final} \geq -0.8$$
$$\text{Minority consensus: } M_{final} < -0.8$$

Note that our definition of consensus requires at least 90% of the individuals adopt either of the two opinions (e.g., when $n_+(t) = 0.9 \ and \ n_-(t) = 0.1$ then $M(t) = 0.9 - 0.1 = 0.8$).

- Link densities $l_{+-}(t), l_{++}(t)$ and $l_{--}(t)$ denote the average number of $+-, ++$ and $--$ links for any given node in the network respectively. Note that a $+-$ link is equivalent to a $-+$ link since the links in the model are undirected.

## 3. Simulations

*Codes are available in [GitHub](#).*

In every iteration, the vector $x$ evolves as a Markov process:
$$x(t+dt) = x(t) + \frac{u(t)}{N}$$

Where $u(t)$ is the displacement vector corresponding to the one of the 22 possible reactions in the interaction **(Table T3, T4: supplementary)**. Hence, the evolution of the network is encoded by the ODE – *Equation [1]*:

$$\frac{dx_q(t)}{dt} = \sum_{r=1}^{22} \frac{u_q^r(t) * p^r(t)}{N}$$

Where $q \in \{1,2,3,4\}$ is the component index of $x$, and $p^r$ is the probability of the $r^{th}$ reaction where $r \in \{1,2,3,...,22\}$ **(Table T3, T4: supplementary)**.

Starting from the initial configuration, we obtain numerical simulations by performing RK-4 integration on equation [1] using the *ode* function in the R package *deSolve*[56].

Unless stated otherwise, we used an initial configuration:

$$n_+(0) = 0.8 \; ; \; n_-(0) = 0.2 \Rightarrow M(0) = 0.6$$

$$l_{++}(0) = 5 * n_+(0) = 4$$
$$l_{--}(0) = 5 * n_-(0) = 1$$
$$l_{+-} = \frac{5}{2} = 2.5$$

To check for robustness over different initial conditions, we varied the majority to minority ratio to 60/40 (weak majority) and 50/50 (no majority) in the supplement.

The simulations were run for $T = 10,000$ time steps with population size of $N = 100$.

# Results

## 1. High stubbornness dominates decision-making

First, we investigated how preferences affect group-level outcomes in the absence of link updating. We examined how the final state of magnetization ($M_{final}$) varies as a function of the stubbornness ($\beta_+$ and $\beta_-$) of the two subpopulations. Increasing the stubbornness of a subpopulation biases the final state in favor of their preference **(Figure 3)**. Stalemates occurred in a region where the products of stubbornness and initial population sizes (net preference) for each subpopulation were similar ($\alpha \approx 1$; grey area). When the majority had a stronger net preference than the minority ($\alpha > 1$), it dominated the final state. Additionally, majority consensus ($M_{final} > 0.8$) occurred when the minority stubbornness was low. When the minority had a stronger net preference ($\alpha < 1$), in turn, it could dominate the final state. However, minority consensus ($M_{final} < -0.8$) never occurred in static networks. Finally, when both subpopulations were highly stubborn, all final states were non-consensus. Overall, being stubborn can shape the consensus outcome in one's favor, but high stubbornness in both subpopulations impedes reaching any consensus.

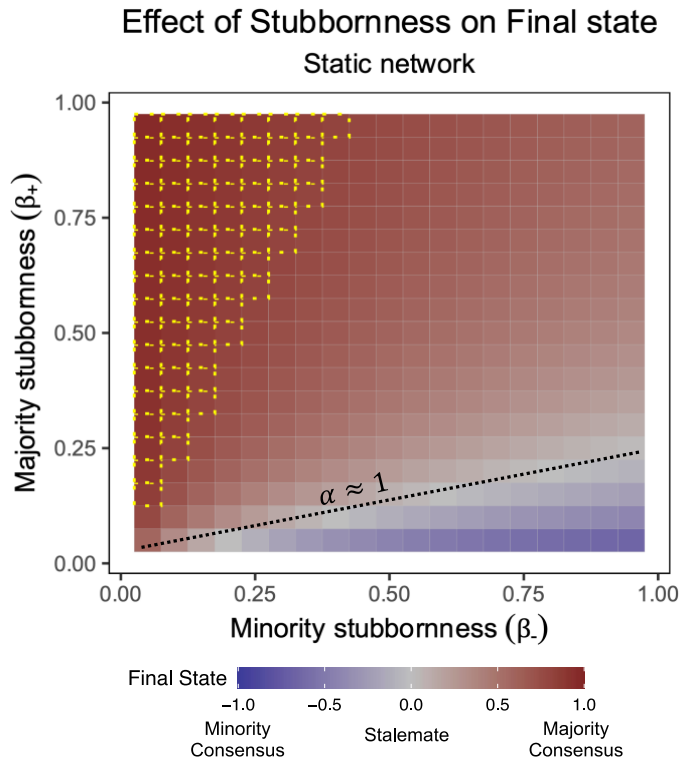

**Figure 3:** Final state ($M_f$) as a function of majority ($\beta_+$) and minority stubbornness ($\beta_-$), for the Incoming model (Outgoing model showed similar behavior (see **Figure S1**)). Final states were biased towards the subpopulation that had a higher net preference. Yellow dotted lines indicate the region where a consensus is reached. Black dotted line indicates the region with a balance of preferences such that stalemates occurred. Each square represents one simulation with the given parameter values for stubbornness. Here, no link updating is included, i.e.,($\phi_1 = 0, \phi_2 = 0$).

## 2. Differences in disagreement avoidance drive consensus outcomes

Link updating strategies are likely to have the strongest impact on group level outcomes when the difference between net preferences of the two subpopulations are small. Therefore, we began by analyzing how the final state of Magnetisation ($M_{final}$) varies as a function of disagreement avoidance ($\phi_2$) and agreement avoidance ($\phi_1$) for similar net preferences ($\alpha \approx 1$). Link updating probabilities were also varied independently for the two subpopulations to study how link updating could help a subpopulation dominate the outcome. We found that, for the Incoming Model (where links are updated by individuals who copy their neighbours), the final state was biased towards the preference of the disagreement avoiding subpopulation **(Figure 4A)**. In other words, when individuals controlled who they received information from, they were more likely to sway the group if they tended to break links with disagreeing others. Conversely, for the Outgoing model (where links are updated by individuals who convince their neighbours), the final state was biased away from the preference of the disagreement avoiding subpopulation **(Figure 4A)**. In other words, when individuals controlled who they sent information to, the subpopulation that retained links with disagreeing others dominated the decision-making outcome.

The size of the bias due to agreement avoidance was negligible in contrast to bias due to disagreement avoidance. In comparison to disagreement avoidance, the bias was reversed for agreement avoidance: for the incoming model, the final state was biased towards the preference of the subpopulation that was less likely to break links with others who agreed. Conversely, for the outgoing model, the final state was biased towards the preference of the subpopulation more likely to break links to others who agreed. Furthermore, the final state remained a stalemate when both subpopulations adopted the same link updating strategies. Overall, these results show that if the overall net preferences are similar between subpopulations, when different subpopulations employ different link updating strategies, it can strongly bias the resulting group-level outcome.

To further examine how link updating strategies affect consensus outcomes, we varied the probability of disagreement avoidance ($\phi_2$) in both subpopulations. Remarkably, differences in disagreement avoidance could change the final state from stalemate to consensus in either direction. In the Incoming Model, consensus decisions were biased toward the preference of the subpopulation with a higher probability of breaking links with disagreeing others **(Figure 4B)**. In contrast, for the Outgoing Model, the subpopulation with a lower probability to break disagreeing links was favoured **(Figure 4C)**. Importantly, no consensus occurred when both subpopulations avoided disagreements to a similar extent, indicating that it is the difference in disagreement avoidance that shapes consensus outcomes. Crucially, these results held true for both majority and minority subpopulations. Thus, a sufficiently stubborn minority can achieve consensus when they have relatively higher probability of breaking links compared to majority individuals while receiving opinions, or a lower probability of breaking links compared to majority individuals while sending opinions.

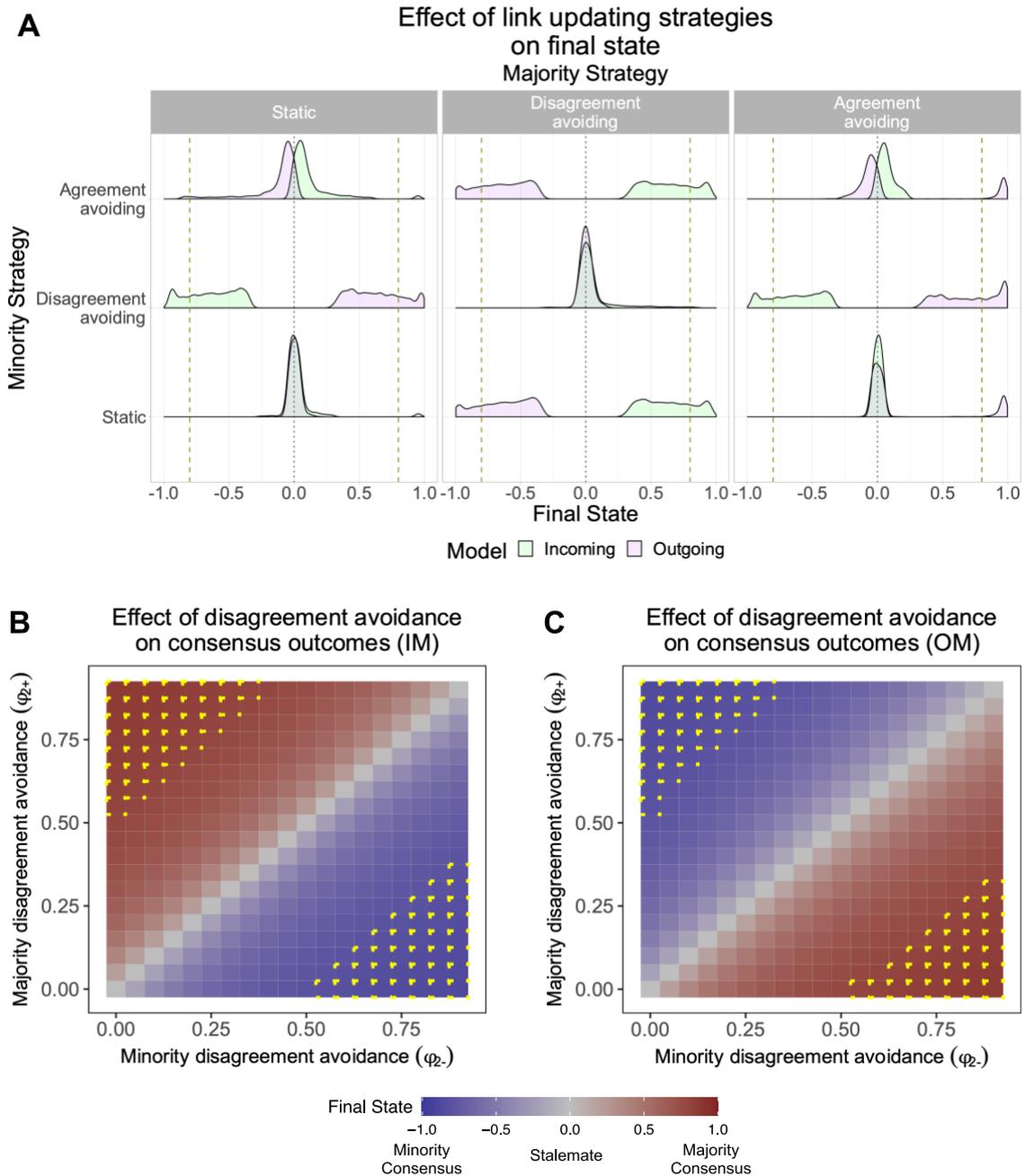

**Figure 4:** Final state ($M_f$) as a function of link updating strategies, for Incoming and Outgoing models. Subpopulations have similar net preferences, $\alpha \approx 1$.

**(A)** Decision outcomes are biased towards the subpopulation that avoids disagreements more in incoming model (green) and less in outgoing model (purple); The plots are obtained for 1500 random points with $\beta_+ \in [0, 0.2]$ and $\beta_- \in [0, 0.8]$; The points indicate distribution of final states for the randomized points. Black dotted lines indicate stalemates and yellow dashed lines indicate consensus.
Strategies: None ($\phi_1 = 0, \phi_2 = 0$); Disagreement avoidance ($\phi_1 = 0, \phi_2 = 0.5$); and Agreement avoidance ($\phi_1 = 0.5, \phi_2 = 0$).

**(B,C)** Disagreement avoidance can bias decision outcomes to stalemates or consensus favoring either subpopulation. Yellow lines indicate regions where consensus is reached. Stubbornness: ($\beta_+ = 0.05, \beta_- = 0.20$).

Beyond that, analyses outside of the region with similar net preferences show that link updating strategies led to consensus outcomes only when the entire population was not highly stubborn **(Figures S1 and S2)**. Thus, consensus outcomes are facilitated by an interplay of differential disagreement avoidance and weakly or similarly stubborn factions.

## 3. Decision-making speed

Finally, we studied the variation in convergence speed (i.e., speed of consensus) with different link and opinion updating properties. We defined convergence speed as the inverse of the time taken to converge to either majority or minority consensus $\left(S_c = \frac{1}{T_c}\right)$. Firstly, we investigated how the speed to majority consensus varies as a function of majority ($\phi_{1+}, \phi_{2+}$) and minority ($\phi_{1-}, \phi_{2-}$) link updating strategies. Speed to majority consensus was consistently related with the direction of outcome bias due to disagreement avoidance by the majority or minority **(Figure 5A)**. As the majority disagreement avoidance increased ($\phi_{2+}$), majority consensus sped up for the Incoming model but slowed down for the Outgoing model. The effect was exactly reversed for the minority: as minority disagreement avoidance ($\phi_{2-}$) increased, majority consensus slowed down for the Incoming model and sped up for the Outgoing model. Similar results were found for cases with minority consensus **(Figure S3)**. Agreement avoidance ($\phi_{1+}, \phi_{1-}$), in contrast, had a negligible influence on decision-making speeds **(Figure 5B)**, paralleling its weak influence over decision outcomes.

Next, we studied how convergence speed varied as a function of stubbornness ($\beta_+, \beta_-$) and inertia ($\lambda$). Importantly, for low values of $\lambda$, inertia had a negligible influence on the final state **(Figure S4)**; the influence of inertia on group level outcomes was prominent only at high values of inertia and disagreement avoidance. In the incoming model, inertia sped up majority consensus formation **(Figure 5C)**. The advantage of increased inertia while receiving opinions can be explained by a reduced rate at which individuals in the majority switched to a minority opinion. Hence, inertia in opinion updating when receiving opinions led to a "slower-is-faster" effect, with increased inertia actually speeding up consensus formation. Conversely, for the outgoing model, inertia led to a lower convergence speed. Increased inertia while sending opinions led to a decreased rate at which the minority adopted the majority opinion, suggesting that exhibiting inertia while sending opinions impedes consensus decision-making. We obtained similar results for minority consensus **(Figure S5A)**, where inertia slowed down the convergence speed to minority consensus for both incoming and outgoing interactions. Finally, we demonstrated how increasing stubbornness ($\beta_+$) of the subpopulation that shapes the consensus outcome (majority) led to faster convergence speeds **(Figure 5D)**. These results held true for both majority and minority consensus **(Figure S5B)**.

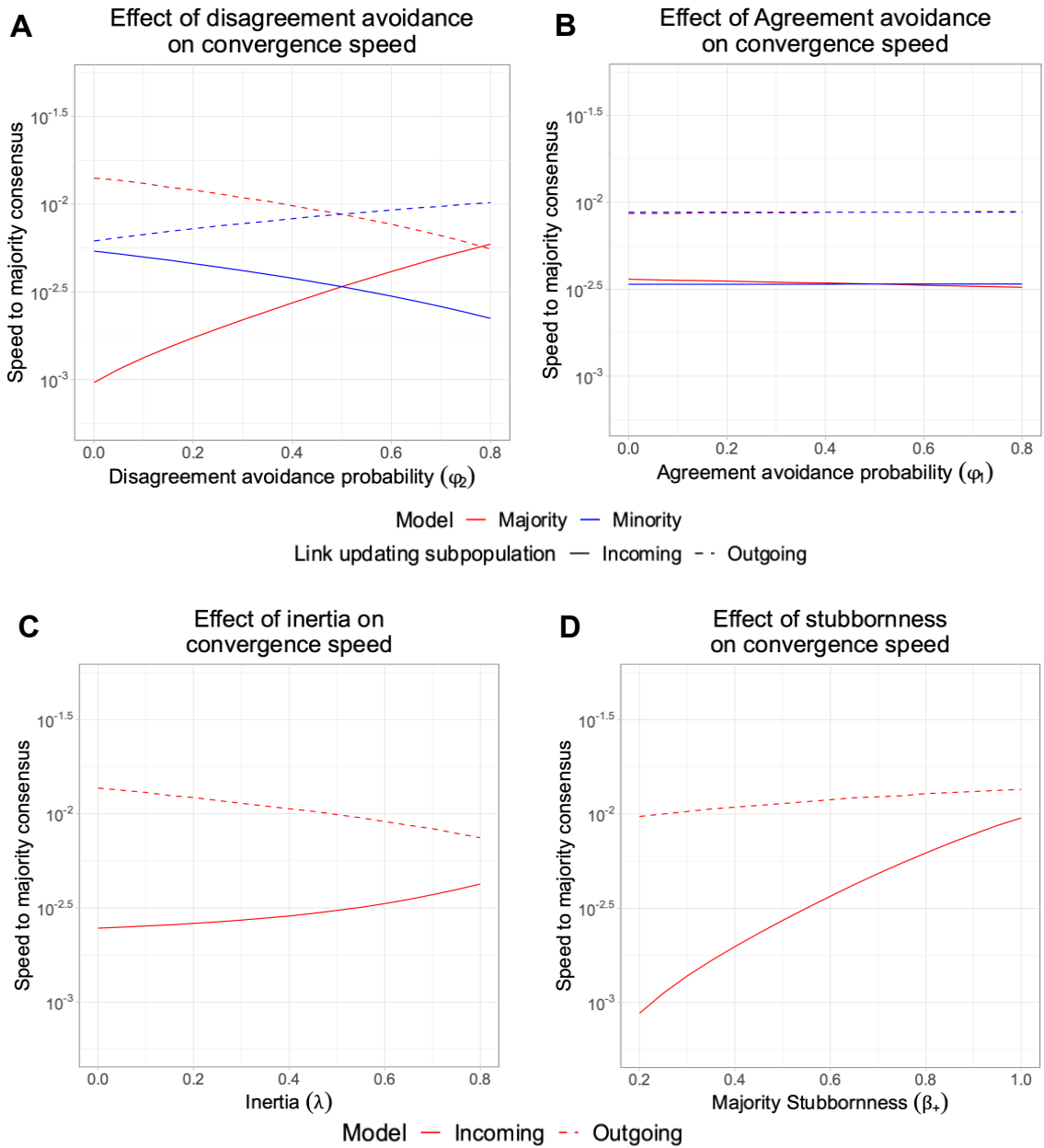

**Figure 5:** Convergence speed to majority consensus as a function of (A,B) link updating and (C,D) opinion updating tendencies.

Convergence Speed (in log scale), $S_c = (\frac{1}{T_c})$ where $T_c$ is time units taken to reach consensus.

**(A):** Majority consensus was reached faster when the majority was more likely to break disagreeing links (higher $\phi_{2+}$) in the incoming model and maintain disagreeing links (lower $\phi_{2+}$) in the outgoing model. Effects were reversed for minority disagreement avoidance ($\phi_{2-}$). **(B):** Effects of agreement avoidance ($\phi_{1+}, \phi_{1-}$) were negligible.
**(C):** Inertia ($\lambda$) speeds up convergence to majority consensus for the incoming model, whereas it slows down convergence speed for the outgoing model. (For higher values of inertia ($\lambda > 0.8$), no consensus occurred). **(D):** Increasing majority stubbornness ($\beta_+$) speeds up convergence to majority consensus for both directions of communication.

# Discussion

Our results demonstrate that link updating strategies could be decisive in determining who dominates collective decision-making outcomes. Group level outcomes were biased in favor of the subpopulation that broke links with disagreeing others when receiving opinions or retained links with disagreeing others when sending opinions. Crucially, we highlight that the consequences of link updating are dependent on the direction of communication, that is, whether individuals choose who to receive opinions from or who to send opinions to. The tendency to break links with disagreeing others could also determine whether the group reached consensus or became locked in a stalemate. However, the effect of link updating was diluted when either of the subpopulations had a much stronger preference than the other. Surprisingly, the propensity to break or retain links with agreeing others had a negligible effect on decision-making speeds and group level outcomes.

Social influence is exerted through communication links and hence, strategies that determine how these links are formed are key to negotiating collective outcomes. We show that strategic link updating can help minorities overturn majorities and maintain dominance. This finding is consistent with multiple studies[57–59] about how homophily and heterophily influence degree rankings and norm adoptions, and potentially help intervene on structural inequality. Avoiding links with disagreeing others promotes homophily (tendency to seek out like-minded others) and leads individuals to seek out and assort into clusters with similar opinionated others. As previously shown[44], homophilous tendencies can give rise to polarized networks and reduced cross talk between dissimilar minded individuals. Assorting into clusters of similar minded others is detrimental when convincing others to adopt an opinion that they disagree with. Hence, retaining outgoing links with disagreeing others is crucial to dominate consensus outcomes. In contrast, clustering increases communication between similar minded others and reinforces their opinion. We showed that this reinforcement can especially help to dominate consensus outcomes when the link updating population decides which opinions to receive, which is in line with past research that found social reinforcement due to clustering[41]. Additionally, we showed that biases in the global outcomes due to disagreement avoidance also show parallel effects on global convergence speeds. This finding is similar to findings that homophily correlates with convergence speeds differently for models that differentiate opinion copying and convincing behavior in interactions[52]. Thus, we demonstrate how individuals can manipulate their links with disagreeing others to dominate consensus outcomes.

We highlight that biases due to link updating strongly interplay with an important nature of the link updating autonomy: namely, whether individuals choose who to receive information from or who send information to. Classical studies around echo chambers have concentrated primarily on information consumption[21,24], where individuals tend to stabilize their opinions by choosing to receive self-reinforcing information. We show that while such social reinforcement can help dominate collective outcomes while receiving opinions, communication with unlike-minded others is crucial in dominating outcomes while sending opinions. This crucial yet overlooked feature of who has the autonomy of updating links is often dependent on the link selection rules of the platform. Thus, we argue that the rules of link selection must be central to studying how communication structures interact with user choice.

Link updating influences network structures, and network structures influence decision outcomes. Most studies have focused on how link updating shapes network structures[47,58,59] or how network structures shapes decision outcomes[16,19,40], but seldom both. By mapping link updating to decision outcomes, we have shown how agents can select links to manipulate collective outcomes. However, by not observing the topological consequences of link updating, we omit potential insights on how link updating could change the final network structure. Although we find that agreement avoidance has negligible effects on consensus outcomes and speeds, studies on how heterophily influences network topologies[47,59] suggest that agreement avoidance might have consequences on final link structure that do not effect decision-making outcomes or speeds.

We found that highly stubborn individuals in the population adopted opinions favoring their preference, leading to non-consensus outcomes. The final outcome was dominated by the subpopulation that had a higher net preference (product of stubbornness and fraction of individuals with the preference). This finding is consistent with previous literature on how stubborn minorities and zealots can dominate decision-making and lead to undemocratic outcomes[1,19,60]. We emphasize that while highly stubborn individuals can shape outcomes towards their preferences, they risk preventing global consensus from being reached[19,32]. On the other hand, weakly stubborn individuals facilitate consensus outcomes, but risk losing dominance over the outcome[35]. But more importantly, we show that the biases due to link updating only exist under the condition that the difference in the net preferences of the subpopulations is negligible. Outside this regime, stubbornness resulted in individuals adopting the dominant preference as their opinion. Thus, any interventions on link structures would be washed over by the overarching effects of higher net preference of one subpopulation. Our results highlight that while link updating is an important mechanism to bias collective outcomes, these biases are overwhelmed by strong differences in net preference across subpopulations.

Generally, inertia (i.e., a tendency to retain the same opinion over time) slowed down decision-making, corroborating earlier findings on how inertia can lead to delays in behavior spread[39]. However, we also reveal the somewhat counter-intuitive result that in some cases inertia can speed up consensus. In particular, we find that adding inertia leads to a faster convergence to majority consensus for the Incoming model. The mechanism behind this effect is likely that inertia reduces the chances of majority changing their opinion due to social influence from the minority; thereby letting them maintain dominance while still receiving opinions from the minority. While this finding is in line with the "slower-is-faster" effect demonstrated in previous studies[38], we contradict the idea that heterogeneity in inertia among individuals is necessary for this effect.

Our model has a number of limitations. Firstly, due to the mean-field approximation, we don't assign individual identities to agents and assume their behavior to be uniform across all other agents of the same opinion. As a result, we were unable to incorporate microscopic characteristics that motivate individuals to retain or break links such as link utilities[61], rewiring costs and individual states[62]. The mean-field approach also necessitates us to assume no pre-existing network structures or modular structures. As studies have shown previously[63], certain individuals interact more frequently than others due to modular structures arising from social hierarchies, and consensus occurs by discourse among only a few leaders who represent each module. Investigating the effect of modular structures and spatial heterogeneity in our models would be a worthwhile avenue to verify the robustness of our results. Moreover, although our model simulates simple contagion-like dynamics, it cannot be generalized to complex quorum-like dynamics where agents use available information and adopt opinions only when sufficiently many in the group conform[64]. Finally, we depict a scenario where the opinion an agent adopts defines its link updating behavior. Hence, the results may not generalize to situations where the a priori individual behavior shows high variability[59].

By observing the co-evolution of opinion and link dynamics, we examine the effect of low-level interaction processes – disagreement avoidance, agreement avoidance, opinion reset to preference and inertia in opinion update. Our model shows that the communication direction of link updating agents, in terms of sending or receiving opinions, can be a crucial factor that interplays with link updating behavior. When links determine who individuals send opinions to and convince, retaining disagreements can help reduce polarization and prevent stalemate outcomes. On the other hand, though avoiding disagreements may lead to cluster formation and echo chambers, contrary to popular notion[24,26,65] it can actually foster consensus when individuals receive opinions and copy each other. Thus, any stewardship of communication structures in OSNs must be sensitive the context-specific rules of link selection that determine who (receiver or sender) controls the link structures and consequently bias the information flow. Moreover, we establish a boundary condition where net preferences for both subpopulations were similar, outside which stubborn agents dominate decision-making and cause stalemates. Hence, we suggest that interventions on link structures must ensure that the stubbornness of the population is conducive to the intervention.

To conclude, we provide valuable insights into consensus formation in social networks as a group-level outcome of differential link updating strategies and weakly stubborn individuals. We show how breaking links with disagreeing others can be favorable while copying others' opinions but unfavorable while convincing others to adopt your opinion. Importantly, we highlight that the effect of link updating is a function of whether individuals have control over who they receive information from or send information to. Hence, platform-specific rules of who (receiver or sender) gets to control link structures can be decisive in decoding the complex role of link updating in shaping consensus decisions in online social networks.

## Code availability

Codes to generate the simulations and figures can be obtained at:
https://github.com/sharajk/Link_updating.git

## Conflict of interest

The authors declare no conflict of interest

## Acknowledgments

This work was supported by the Max Planck Institute of Animal Behavior and funded by the Deutsche Forschungsgemeinschaft (DFG, German Research Foundation) under Germany's Excellence Strategy – EXC 2117 – 422037984. SK acknowledges funding from the Kishore Vaigyanik Protsahan Yojana (KVPY) fellowship. ASP acknowledges funding from the Gips-Schüle Stiftung. SS acknowledges funding from a startup research grant from IISc Bangalore. We thank Alison Ashbury for helpful comments on the manuscript, as well as the CoCoMo group and CSB lab for useful feedback.

## Author contributions

SK, ASP, NG and HG conceived the study. SK and NG developed and implemented the model. SK and NG analyzed and interpreted the results with inputs from NG, HG, ASP, SS and MKJ. PM reviewed all codes used in the manuscript. SK wrote the first draft of the manuscript. All authors provided feedback on the manuscript and approved the final version.

# Supplementary Material

# Link updating strategies influence consensus decisions as a function of the direction of communication

Sharaj Kunjar*[1,2], Ariana Strandburg-Peshkin[1,3,4], Helge Giese[4,5,6], Pranav Minasandra[1,4,7], Sumantra Sarkar[8], Mohit Kumar Jolly[9], Nico Gradwohl[1,4,5]

[1] Department for the Ecology of Animal Societies, Max Planck Institute of Animal Behavior, Konstanz, Germany
[2] Undergraduate Programme, Indian Institute of Science, Bangalore, India
[3] Department of Biology, University of Konstanz, Germany
[4] Centre for the Advanced Study of Collective Behaviour, University of Konstanz, Germany
[5] Department of Psychology, University of Konstanz, Germany
[6] Heisenberg Chair for Medical Risk Literacy and Evidence-based Decisions, Charité – Universitätsmedizin Berlin, Berlin, Germany
[7] International Max Planck Research School for Quantitative Behavior, Ecology and Evolution, Radolfzell, Germany
[8] Department of Physics, Indian Institute of Science, Bangalore, India
[9] Centre for BioSystems Science and Engineering, Indian Institute of Science, Bangalore, India

*Author to whom correspondence should be addressed to: S Kunjar (sharaj.kunjar@gmail.com)

## Significance Statement

In an age characterized by unprecedented access to information through online social networks, understanding how collective opinions are influenced by users choosing their communication partners is essential to disentangle the mechanisms of polarization and deadlocks. Importantly, the direction of user choice is context dependent: for example, on Twitter, users choose who they receive information from while following public accounts but choose who they send information to while setting Twitter circles. Similarly, on YouTube, users choose who they receive content from when they subscribe to a channel, but creators choose who to send content to while targeting video campaigns. Here, we study how the autonomy that users have in manipulating their communication networks interacts with collective opinions. Using numerical simulations, we show that consensus decisions can be biased by users avoiding disagreeing others (e.g., homophily) as a function of whether they send or receive information. Avoiding disagreeing others assorts users into like-minded clusters, and such clusters facilitated consensus outcomes in favor of users who either retained disagreeing links while sending information or broke disagreeing links while receiving information. Besides highlighting the importance of context-specific effects of user link selection, we show boundary conditions under which these effects matter. These findings shed light on the importance of considering how user choice interacts with communication networks when informing interventions against ideological and structural polarization in media.

# Tables

| Variable | Definition |
|---|---|
| $n_s(t)$ | The fraction of population with opinion 's' at time t, where $s \in \{+, -\}$ <br> $n_+(t) + n_-(t) = 1$ |
| $M(t)$ | Difference in the fraction of populations with + and – opinions <br> $M(t) = n_+(t) - n_-(t)$ |
| $l_{s_1 s_2}$ | Average number of links connecting two nodes with opinions $s_1$ and $s_2$ respectively, where $s_1, s_2 \in \{-, +\}$ |

**Table T1:** Definition of model state variables

| Parameter | Definition |
|---|---|
| $\phi_{1+}, \phi_{1-}$ | Probability of breaking links when both interacting individuals hold the same opinion (agreement avoidance) |
| $\phi_{2+}, \phi_{2-}$ | Probability of breaking links when the interacting individuals hold different opinions (disagreement avoidance) |
| $\beta_+, \beta_-$ | Probability that an individual sticks to or reverts to their stable preference <br> (stubbornness) |
| $\lambda$ | Probability that an individual maintains status quo by holding onto their current opinion (inertia) |

**Table T2:** Definition of model parameters

| $r$ | Initial State | Process | Final State | Vector $(u^r)$ | Probability $(p^r)$ |
|---|---|---|---|---|---|
| 1 | $++$ | Random Rewiring | $+-$ | $(0,1,-1,0)$ | $\widetilde{B} * \phi_{1+} * n_+ n_- * \frac{k_{++}}{k_+}$ |
| 2 | | | $++$ | $(0,0,0,0)$ | $\widetilde{B} * \phi_{1+} * n_+^2 * \frac{k_{++}}{k_+}$ |
| 3 | | No update | $++$ | $(0,0,0,0)$ | $\widetilde{B} * (1-\phi_{1+}) * n_+ * \frac{k_{++}}{k_+}$ |
| 4 | | Preference Switching | $++$ | $(0,0,0,0)$ | $B_+ * n_+ * \frac{k_{++}}{k_+}$ |
| 5 | | | $-+$ | $(-2, k_{++} - k_{+-}, -k_{++}, k_{+-})$ | $B_- * n_+ * \frac{k_{++}}{k_+}$ |
| 6 | $+-$ | Random Rewiring | $++$ | $(0,-1,1,0)$ | $\widetilde{B} * \phi_{2+} * n_+^2 * \frac{k_{+-}}{k_+}$ |
| 7 | | | $+-$ | $(0,0,0,0)$ | $\widetilde{B} * \phi_{2+} * n_+ n_- * \frac{k_{+-}}{k_+}$ |
| 8 | | Spin Update | $--$ | $(-2, k_{++} - k_{+-}, -k_{++}, k_{+-})$ | $\widetilde{B} * (1-\phi_{2+}) * (1-\lambda) * n_+ * \frac{k_{+-}}{k_+}$ |
| 9 | | No update | $+-$ | $(0,0,0,0)$ | $\widetilde{B} * (1-\phi_{2+}) * \lambda * n_+ * \frac{k_{+-}}{k_+}$ |
| 10 | | Preference Switching | $+-$ | $(0,0,0,0)$ | $B_+ * n_+ * \frac{k_{+-}}{k_+}$ |
| 11 | | | $--$ | $(-2, k_{++} - k_{+-}, -k_{++}, k_{+-})$ | $B_- * n_+ * \frac{k_{+-}}{k_+}$ |
| 12 | $-+$ | Random Rewiring | $--$ | $(0,-1,0,1)$ | $\widetilde{B} * \phi_{2-} * n_-^2 * \frac{k_{-+}}{k_-}$ |
| 13 | | | $-+$ | $(0,0,0,0)$ | $\widetilde{B} * \phi_{2-} * n_+ n_- * \frac{k_{-+}}{k_-}$ |
| 14 | | Spin Update | $++$ | $(2, k_{--} - k_{-+}, k_{-+}, -k_{--})$ | $\widetilde{B} * (1-\phi_{2-}) * (1-\lambda) * n_- * \frac{k_{-+}}{k_-}$ |
| 15 | | No update | $-+$ | $(0,0,0,0)$ | $\widetilde{B} * (1-\phi_{2-}) * \lambda * n_- * \frac{k_{-+}}{k_-}$ |
| 16 | | Preference Switching | $-+$ | $(0,0,0,0)$ | $B_+ * n_- * \frac{k_{-+}}{k_-}$ |
| 17 | | | $++$ | $(2, k_{--} - k_{-+}, k_{-+}, -k_{--})$ | $B_- * n_- * \frac{k_{-+}}{k_-}$ |
| 18 | $--$ | Random Rewiring | $-+$ | $(0,1,0,-1)$ | $\widetilde{B} * \phi_{1-} * n_+ n_- * \frac{k_{--}}{k_-}$ |
| 19 | | | $--$ | $(0,0,0,0)$ | $\widetilde{B} * \phi_{1-} * n_-^2 * \frac{k_{--}}{k_-}$ |
| 20 | | No Update | $--$ | $(0,0,0,0)$ | $\widetilde{B} * (1-\phi_{1-}) * n_- * \frac{k_{--}}{k_-}$ |
| 21 | | Preference Switching | $--$ | $(0,0,0,0)$ | $B_+ * n_- * \frac{k_{--}}{k_-}$ |
| 22 | | | $+-$ | $(2, k_{--} - k_{-+}, k_{-+}, -k_{--})$ | $B_- * n_- * \frac{k_{--}}{k_-}$ |

**Table T3:** List of reactions when two nodes interact, with displacement vector and probability of the reaction; for Incoming Model; $\beta_+ * n_+(0) \equiv B_+$ ; $\beta_- * n_-(0) \equiv B_-$ ; $\{1 - \beta_+ * n_+(0) - \beta_- * n_-(0)\} \equiv \widetilde{B}$ ; $k_{++} = \frac{l_{++}}{n_+}$ ; $k_{+-} = \frac{l_{+-}}{n_+}$ ; $k_{-+} = \frac{l_{+-}}{n_-}$ ; $k_{--} = \frac{l_{--}}{n_-}$

| $r$ | Initial State | Process | Final State | Vector ($u^r$) | Probability ($p^r$) |
|---|---|---|---|---|---|
| 1 | $++$ | Random Rewiring | $+-$ | $(0,1,-1,0)$ | $\widetilde{B} * \phi_{1+} * n_+ n_- * \frac{k_{++}}{k_+}$ |
| 2 | | Random Rewiring | $++$ | $(0,0,0,0)$ | $\widetilde{B} * \phi_{1+} * n_+^2 * \frac{k_{++}}{k_+}$ |
| 3 | | No update | $++$ | $(0,0,0,0)$ | $\widetilde{B} * (1-\phi_{1+}) * n_+ * \frac{k_{++}}{k_+}$ |
| 4 | | Preference Switching | $++$ | $(0,0,0,0)$ | $B_+ * n_+ * \frac{k_{++}}{k_+}$ |
| 5 | | Preference Switching | $-+$ | $(-2, k_{++}-k_{+-}, -k_{++}, k_{+-})$ | $B_- * n_+ * \frac{k_{++}}{k_+}$ |
| 6 | $+-$ | Random Rewiring | $++$ | $(0,-1,1,0)$ | $\widetilde{B} * \phi_{2+} * n_+^2 * \frac{k_{+-}}{k_+}$ |
| 7 | | Random Rewiring | $+-$ | $(0,0,0,0)$ | $\widetilde{B} * \phi_{2+} * n_+ n_- * \frac{k_{+-}}{k_+}$ |
| 8 | | Spin Update | $--$ | $(2, k_{--}-k_{-+}, k_{-+}, -k_{--})$ | $\widetilde{B} * (1-\phi_{2+})*(1-\lambda) * n_+ * \frac{k_{+-}}{k_+}$ |
| 9 | | No update | $+-$ | $(0,0,0,0)$ | $\widetilde{B} * (1-\phi_{2+}) * \lambda * n_+ * \frac{k_{+-}}{k_+}$ |
| 10 | | Preference Switching | $+-$ | $(0,0,0,0)$ | $B_+ * n_+ * \frac{k_{+-}}{k_+}$ |
| 11 | | Preference Switching | $--$ | $(-2, k_{++}-k_{+-}, -k_{++}, k_{+-})$ | $B_- * n_+ * \frac{k_{+-}}{k_+}$ |
| 12 | $-+$ | Random Rewiring | $--$ | $(0,-1,0,1)$ | $\widetilde{B} * \phi_{2-} * n_-^2 * \frac{k_{-+}}{k_-}$ |
| 13 | | Random Rewiring | $-+$ | $(0,0,0,0)$ | $\widetilde{B} * \phi_{2-} * n_+ n_- * \frac{k_{-+}}{k_-}$ |
| 14 | | Spin Update | $++$ | $(-2, k_{++}-k_{+-}, -k_{++}, k_{+-})$ | $\widetilde{B} * (1-\phi_{2-})*(1-\lambda) * n_- * \frac{k_{-+}}{k_-}$ |
| 15 | | No update | $-+$ | $(0,0,0,0)$ | $\widetilde{B} * (1-\phi_{2-}) * \lambda * n_- * \frac{k_{-+}}{k_-}$ |
| 16 | | Preference Switching | $-+$ | $(0,0,0,0)$ | $B_+ * n_- * \frac{k_{-+}}{k_-}$ |
| 17 | | Preference Switching | $++$ | $(2, k_{--}-k_{-+}, k_{-+}, -k_{--})$ | $B_- * n_- * \frac{k_{-+}}{k_-}$ |
| 18 | $--$ | Random Rewiring | $-+$ | $(0,1,0,-1)$ | $\widetilde{B} * \phi_{1-} * n_+ n_- * \frac{k_{--}}{k_-}$ |
| 19 | | Random Rewiring | $--$ | $(0,0,0,0)$ | $\widetilde{B} * \phi_{1-} * n_-^2 * \frac{k_{--}}{k_-}$ |
| 20 | | No Update | $--$ | $(0,0,0,0)$ | $\widetilde{B} * (1-\phi_{1-}) * n_- * \frac{k_{--}}{k_-}$ |
| 21 | | Preference Switching | $--$ | $(0,0,0,0)$ | $B_+ * n_- * \frac{k_{--}}{k_-}$ |
| 22 | | Preference Switching | $+-$ | $(2, k_{--}-k_{-+}, k_{-+}, -k_{--})$ | $B_- * n_- * \frac{k_{--}}{k_-}$ |

**Table T4:** List of reactions when two nodes interact, with displacement vector and probability of the reaction; for Outgoing Model; $\beta_+ * n_+(0) \equiv B_+$ ; $\beta_- * n_-(0) \equiv B_-$ ; $\{1 - \beta_+ * n_+(0) - \beta_- * n_-(0)\} \equiv \widetilde{B}$ ; $k_{++} = \frac{l_{++}}{n_+}$; $k_{+-} = \frac{l_{+-}}{n_+}$; $k_{-+} = \frac{l_{+-}}{n_-}$; $k_{--} = \frac{l_{--}}{n_-}$

# S1. High stubbornness dominates decision-making

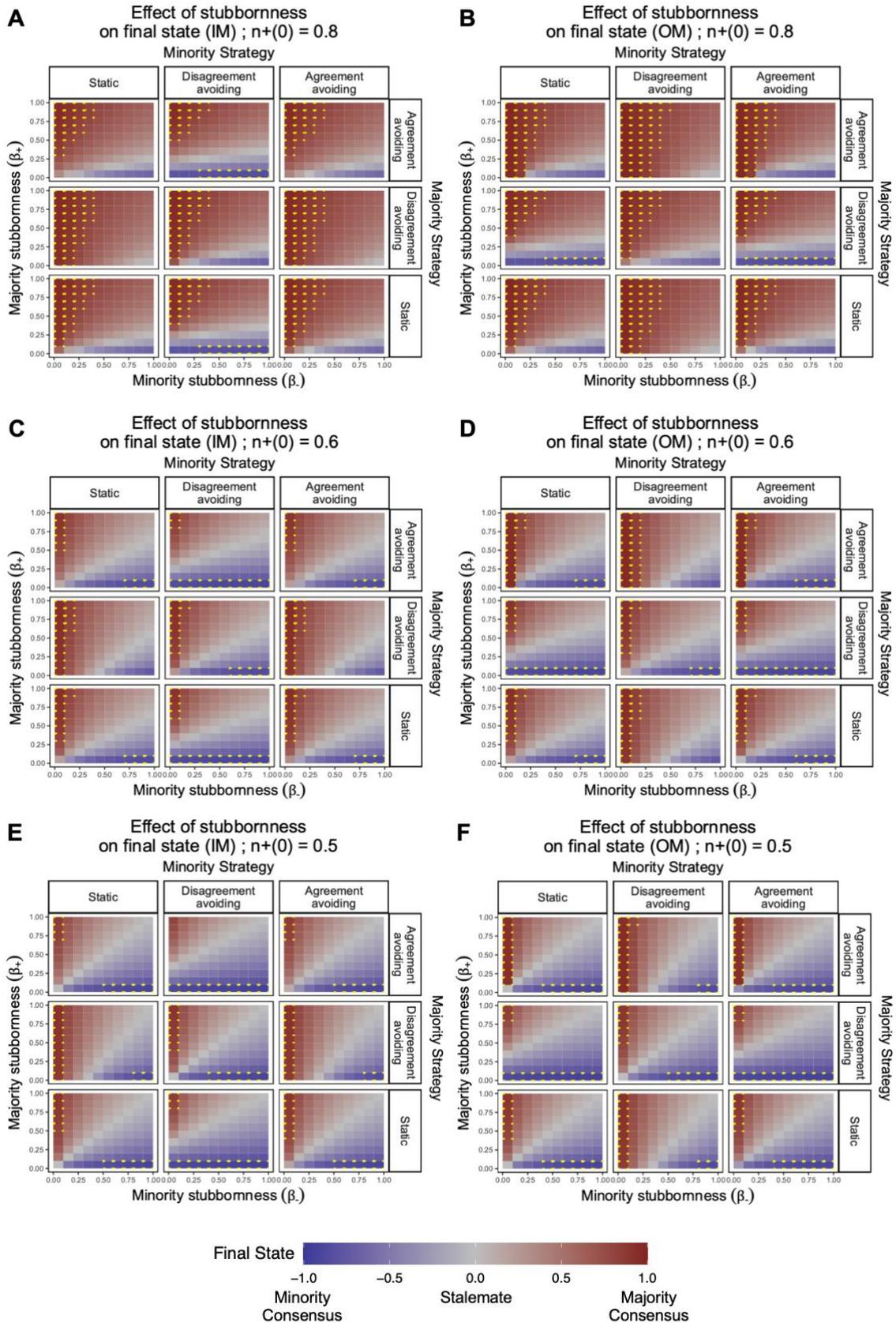

**Figure S1:** Final state of Magnetisation ($M_f$) as a function of majority ($\beta_+$) and minority stubbornness ($\beta_-$) for different link updating strategies (facets). Fraction of initial subpopulation sizes of majority to minority is **(A,B)** 80/20 **(C,D)** 60/40 and **(E,F)** 50/50. Highly stubborn individuals shape the final state towards their preferred outcome. The net preference (product of stubbornness and initial subpopulation size) is equal in the gray region ($n_+(0) * \beta_+ \approx n_-(0) * \beta_-$) where stalemates occur. Disagreement avoidance shifts the gray region, indicating that link updating strategies bias decision outcomes. The slope of the gray region is determined by the ratio of majority and minority subpopulation sizes ($\frac{n_+(0)}{n_-(0)}$). Yellow dotted lines indicate regions where consensus is reached. Strategies: None ($\phi_1 = 0, \phi_2 = 0$); Disagreement avoidance ($\phi_1 = 0, \phi_2 = 0.5$); and Agreement avoidance ($\phi_1 = 0.5, \phi_2 = 0$).

# S2. Differential disagreement avoidance leads to consensus outcomes

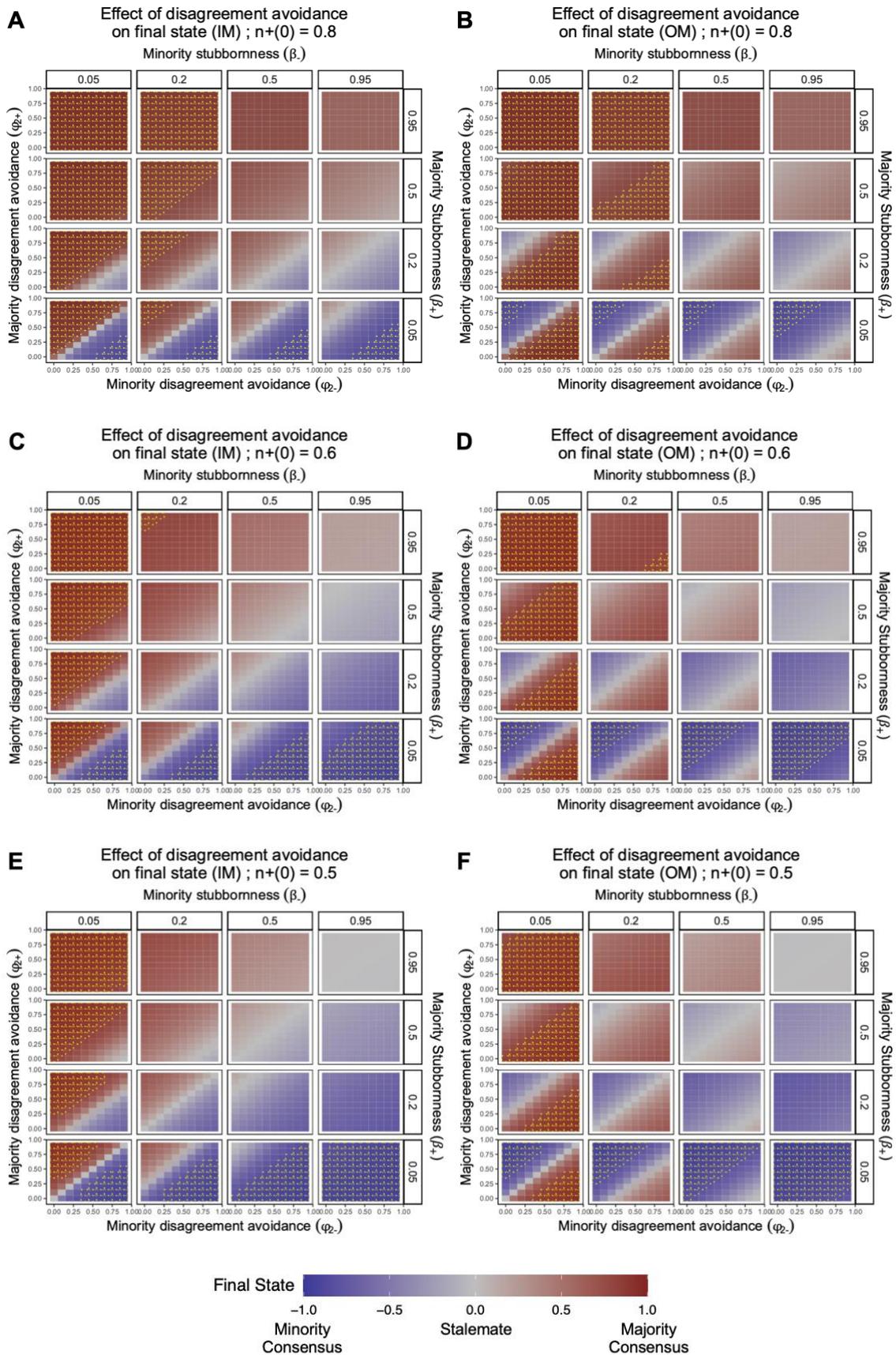

**Figure S2:** Final state as a function of majority ($\phi_{2+}$) and minority ($\phi_{2-}$) disagreement avoidance probabilities, for incoming (left) and outgoing (right) models. Fraction of initial subpopulation sizes of majority to minority is **(A,B)** 80/20 **(C,D)** 60/40 and **(E,F)** 50/50. Facets indicates stubbornness. Yellow lines indicate regions where consensus occurs. Consensus outcomes are facilitated by differential disagreement avoidance and weakly stubborn individuals.

# S3. Decision making rates – Link updating strategies

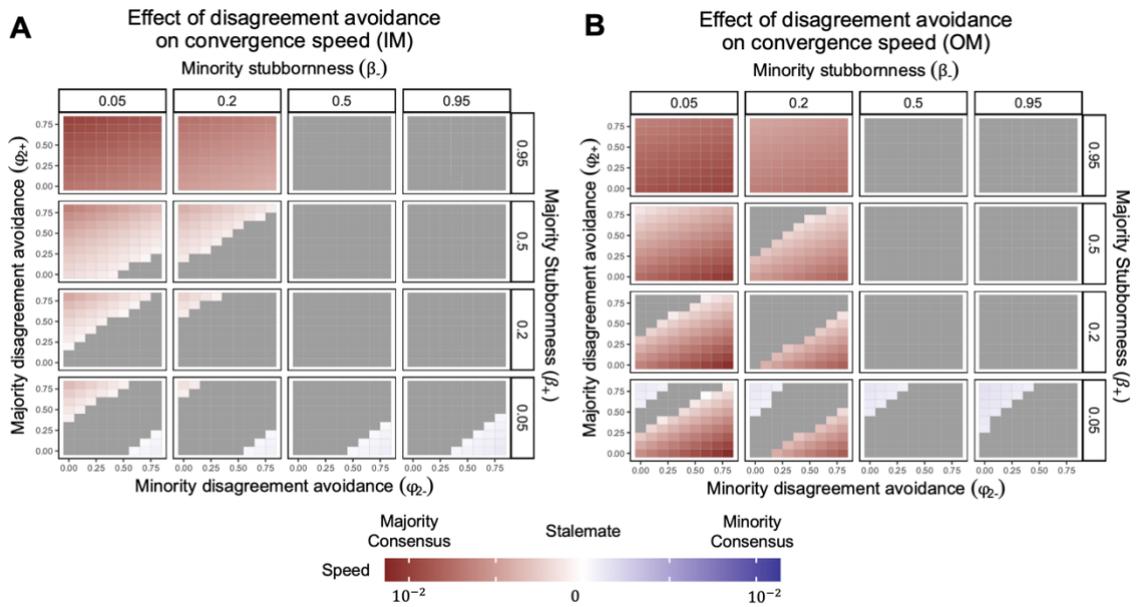

**Figure S3:** Convergence speed as a function of majority ($\phi_{2+}$) and minority ($\phi_{2-}$) disagreement avoidance probabilities, for incoming **(A)** and outgoing **(B)** models. The fraction of initial subpopulation sizes for majority to minority is 80/20. Facets indicates stubbornness of the two subpopulations. Speed is in log scale. Gray regions indicate no consensus outcomes. Biases due to disagreement avoidance on convergence speed parallels with the biases on the outcome.

# S4. Low inertia has negligible impact on collective outcomes

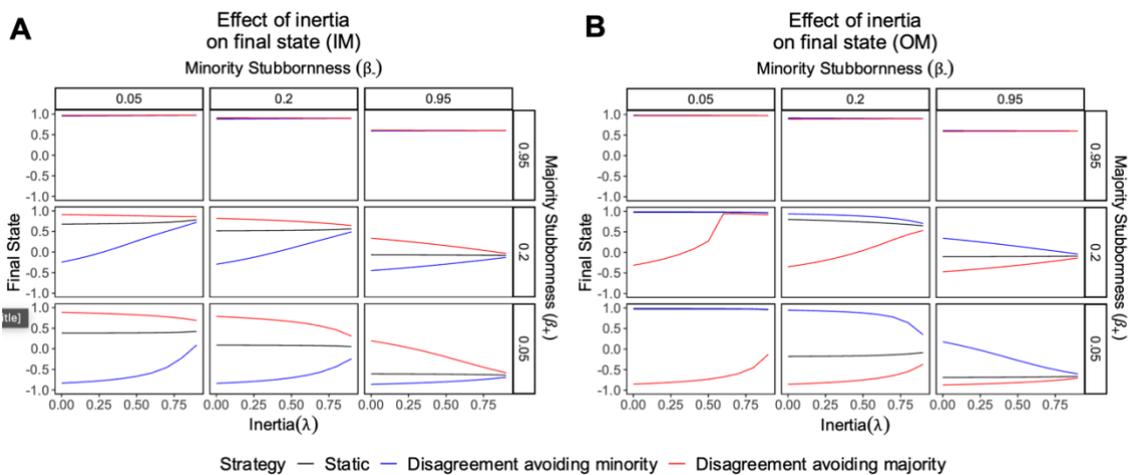

**Figure S4:** Final state ($M_f$) as a function of inertia ($\lambda$), for incoming **(A)** and outgoing **(B)** models. Facets indicate stubbornness. The fraction of initial subpopulation sizes for majority to minority is 80/20. Inertia only effects outcomes at high values and differential disagreement avoidance.

## S5. Decision-making rates – Stubbornness and inertia

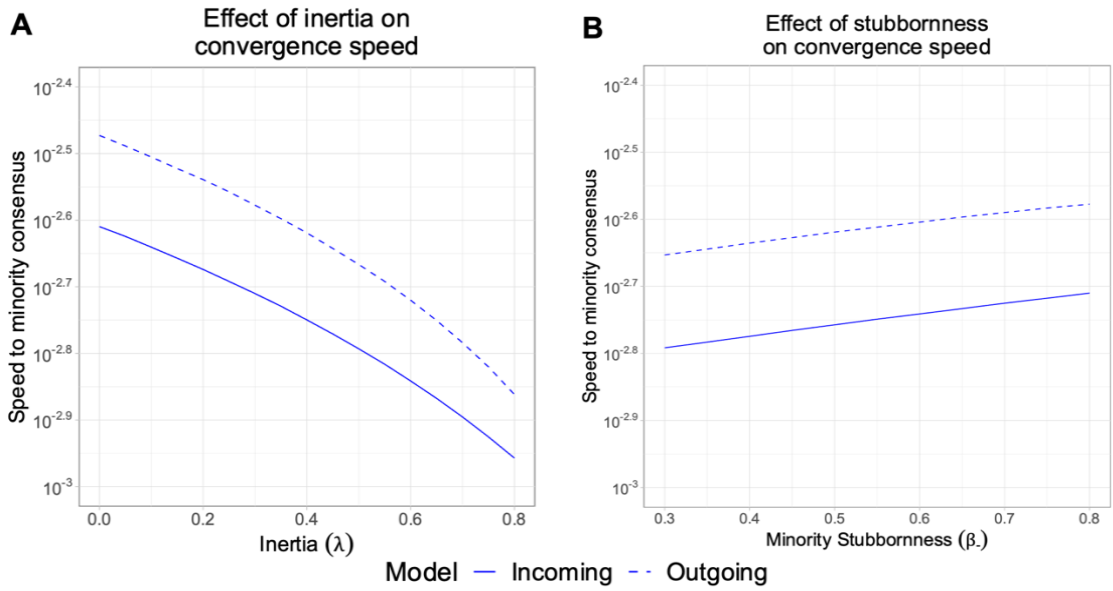

**Figure S5:** Convergence speed to minority consensus as a function of inertia **(A)** and minority stubbornness **(B)**. Convergence speed is positively correlated to stubbornness but negatively correlated to inertia, for both incoming and outgoing models.